\documentclass[prl,twocolumn,showpacs,preprintnumbers,amsmath,amssymb]{revtex4}
\usepackage{graphicx}
\newcommand{\beq}{\begin{equation}}
\newcommand{\eeq}{\end{equation}}
\newcommand{\beqa}{\begin{eqnarray}}
\newcommand{\eeqa}{\end{eqnarray}}

\begin{document}
\preprint{}
\title{Nuclear Saturation with Low Momentum Interactions.}
\author{E.N.E. van Dalen}
\author{H. M\"uther}
\affiliation{Institut
f$\ddot{\textrm{u}}$r Theoretische Physik, Universit$\ddot{\textrm{a}}$t
T$\ddot{\textrm{u}}$bingen,
Auf der Morgenstelle 14, D-72076 T$\ddot{\textrm{u}}$bingen, Germany}
\begin{abstract}
Relativistic effects are investigated in nuclear matter calculations employing
renormalized low-momentum nucleon-nucleon ($NN$) interactions. It is
demonstrated that the relativistic effects cure a problem of non-relativistic
low-momentum interactions, which fail to reproduce saturation of nuclear
matter.  Including relativistic effects, one already obtains saturation in a
Hartree-Fock calculation. Brueckner-Hartree-Fock calculations lead to a further
improvement of the saturation properties. The results are rather insensitive to
the realistic $NN$ interaction on which they are based.
\end{abstract}
\maketitle

A fundamental question in nuclear physics is the role played by relativistic
effects. At first 
sight, only rather moderate relativistic effects are expected, since the
velocity of nucleons in nuclear matter reaches less than one fourth of the light
velocity and the kinetic and potential energies are small as compared to the rest
mass of the nucleons. However, considering a meson-exchange model for the NN
interaction the weak single-particle potential occurs as a sum of a very 
attractive scalar field and a repulsive vector field, each of them of a size as
large as several hundred MeV. This leads to a modification of the Dirac spinors
for the nucleons in the nuclear medium and can be expressed in terms of an
effective Dirac mass, which tends to decrease with increasing
density\cite{walecka,brockm,vandalen:2005}. The matrix elements of the $NN$
interaction in the nuclear medium should be evaluated using these dressed Dirac
spinors and therefore depend on the density of the matter considered. These
relativistic effects lead to a successful microscopic description of the 
saturation properties of nuclear matter in
Dirac-Brueckner-Hartree-Fock (DBHF)
approach\cite{brockm,muether:rel,DBHF} 
without a need to include many-nucleon forces, which are required in
non-relativistic investigations\cite{wiringa}.

Apart from the division between relativistic and non-relativistic approaches, one
can distinguish between investigations which are based on phenomenological
interactions and those which are based on realistic $NN$ interactions.  The
parameters of the former interactions have been adjusted to describe properties
of isospin symmetric nuclear matter and of nuclei in the valley of $\beta$
stability. These phenomenological models, such as Gogny~\cite{gogny},
Skyrme~\cite{sk1}, or relativistic mean field (RMF) approaches~\cite{rmf} provide
a simple description of the mean field in terms of local single-particle
densities. However, a disadvantage is that their predictive power may be rather
limited, in particular for highly isospin asymmetric nuclear matter and nuclei
far away from the line of $\beta$ stability. The study of exactly these nuclei is
of high interest with the forthcoming new generation of radioactive beam
facilities like the future GSI facility FAIR in Germany and the RIA facility
planned in the United States of America.

One the other hand, the microscopic approaches based on realistic $NN$
interactions, like the Bonn interactions\cite{brockm}, have a high predictive
power, which should give one confidence when the model is used in extreme cases
like in a highly isospin asymmetric nuclear environment. These high-precision
free-space $NN$ interactions are adjusted to describe the experimental data of
the $NN$ interaction. However, the strong short range and tensor components of
such realistic interactions make it inevitable to employ non-perturbative
approximation schemes for the solution of the many-body problem.  Therefore,
calculations with such interactions are restricted to very light nuclei due to
the dramatic increase of configurations with an increasing number of nucleons. 
Furthermore, such sophisticated calculations will not become feasible for heavier
nuclei or the nuclear structures in the neutron star crust in the near future.

A possible way out of this problem is to restrict the nuclear structure
calculation to the low momentum components by separating the low momentum and
high momentum components of a realistic $NN$ interaction by means of
renormalization techniques~\cite{bogner:2001,bogner:2005,bozek:2006}. If the
cutoff $\Lambda$ is appropriately chosen, i.e. around $\Lambda=2$ fm$^{-1}$, the
resulting low momentum interaction $V_{lowk}$ will still describe the
experimental data of the $NN$ interaction up to the pion threshold.  Moreover, a
very attractive aspect is that this $V_{lowk}$ interaction turns out to be
independent of the underlying realistic interaction $V$.

Using $V_{lowk}$ in a calculation of nuclear matter, one obtains a binding energy
per nucleon increasing with density in a monotonic
way~\cite{bozek:2006,kuckei:2003,goegelein:2009}, unless three-body forces are
added~\cite{bogner:2005}. Thus, the emergence of a saturation point is prevented
in symmetric nuclear matter. Attempts have been made to cure this problem and
recently Siu et al.\cite{siu:2009} suggested to increase the cutoff, account for
the contribution of $pphh$ ring diagrams and a modification of the NN
interaction, the so-called Brown-Rho scaling\cite{brown:2004}. Increasing the
cutoff parameter leads to the necessity to account for correlation effects
($pphh$ ring diagrams) in a non-perturbative way and thereby one of the nice
features of the low momentum interaction, that they can be treated in a
perturbative way, is lost.

Therefore, in the present work, we want to follow a different route: We  include
the relativistic effects discussed above and evaluate for each density the matrix
elements of the bare $NN$ interactions using the Dirac spinors, which are
appropriate for this density. In this work,  the medium properties of the
nucleons, which are used to dress these Dirac spinors, are obtained from the EoS
presented in Refs.~\cite{DBHF}. From this density-dependent bare interaction we
can than deduce a renormalized $V_{lowk}$ interaction, which will depend on the
density, 
using the standard techniques.

For the construction of a low momentum potential, it is necessary to separate the
low momentum and high momentum components of realistic interactions. In our work,
the unitary-model-operator approach (UMOA) ~\cite{suzuki:1982} is used to
disentangle these parts.  As in all well-known model space techniques, we define
the projection operators $P$, which projects onto the low momentum subspace, and
$Q$, which projects onto the complement of this subspace, the high momentum
subspace. Furthermore, these operators P and Q satisfy, of course,  the usual
relations like $P+Q=1$, $P^2=P$, $Q^2=Q$, and $PQ=0=QP$. The aim of the
unitary-model-operator approach is now to define a unitary transformation $U$ in
such a way that the transformed Hamiltonian does not couple $P$ and $Q$, which
means
\begin{equation}
 QU^{-1}HUP = 0
\end{equation}
has to be fulfilled. It leads to an effective Hamiltonian 
\begin{equation}
H_{eff}=h_0+V_{eff},
\label{eq:heff}
\end{equation} 
which contains  a starting Hamiltonian $h_0$ describing the one-body part of the
two-body system and an effective interaction $V_{eff}$.   This effective
interaction is defined in terms of this unitary transformation as
\begin{equation}
  V_{eff} = U^{-1}(h_0 + V) U - h_0,
\label{eq:veff}
\end{equation}
with $V$ representing the bare $NN$ interaction. 
The unitary operator $U$ can be expressed as
\begin{equation}
  U = (1+ \omega -\omega^\dagger )(1+\omega \omega^\dagger + \omega^\dagger
\omega)^{-1/2}
\label{eq:umoa}
\end{equation}
with an operator $\omega$ fulfilling the relation
$\omega = Q\omega P$ such that $\omega^2 = \omega^{\dagger 2} = 0$.
This operator $\omega$  can be  obtained
by first solving the two-body eigenvalue
equation
\begin{equation}
  (h_0 + V ) |\Phi_k \rangle = E_k | \Phi_k \rangle
\label{eq:eigenvalue}
\end{equation}
and afterwards defining the matrix elements of $\omega$ using the eigenstates
$|\Phi_p \rangle$ having the largest overlap with the low momentum space, the
$P$-space~\cite{bozek:2006}. Next, the effective interaction $V_{eff}$ is
calculated as described in \cite{fuji:2004,bozek:2006}. In this way one obtains
the effective Hamiltonian of Eq.~(\ref{eq:heff}), which contains the effective
interaction $V_{eff}$. The eigenvalues, which are obtained by diagonalizing this
effective Hamiltonian in the $P$-space, are identical to those, which are
obtained in the diagonalization of the original Hamiltonian $H=h_0+V$ in the
complete space. 

Next, this model-space scheme can be applied to the effective two-nucleon problem
by considering for the basis states of the two-nucleon system the states
identified by the relative momentum, its modulus and the corresponding partial
wave. For a given partial wave, the states with a relative momentum smaller than
a cutoff $\Lambda$ are identified as the states of the P-space. Therefore,
applying the technique described above leads to the normal effective interaction
$V_{lowk}$.

In order to address the question which role correlations beyond the simple
mean-field or HF approximation play for the density dependent $V_{lowk}(\rho)$
interaction, the effects of $NN$ correlations are taken into account by means of
the BHF approximation. In the BHF approach, one replaces the bare $NN$
interaction V, or in our case the effective interaction $V_{eff}$, used in the HF
approximation by the $G$-matrix which obeys the Bethe-Goldstone equation.  In the
two-particle center of mass frame,  it takes for an effective interaction
$V_{eff}$ with cutoff $\Lambda$ the form
\begin{eqnarray}
G(k',k,\epsilon_k)_{eff}=V_{eff}(k',k)+\int_0^{\Lambda} q^2 dq V_{eff}(k',q)
\nonumber \\ \frac{Q_P}{2\epsilon_k-2 \epsilon_q + i \eta} G(q,k,\epsilon_k),
\label{eq:BG}
\end{eqnarray}  
where $\epsilon_i$ with $i=k,q$ are single-particle energies. Furthermore, the
Pauli operator $Q_P$ prevents scattering to occupied states and, therefore,
restricts the intermediate states to particle states with momenta $q$ which are
above the Fermi energy. In Eq. (~\ref{eq:BG}), we also take into account that 
$V_{eff}$ is designed for a model space with relative momenta smaller than
$\Lambda$. Therefore, the integral in Eq. (~\ref{eq:BG}) is restricted to momenta
$q$ below the cutoff $\Lambda$. In contrast, there is no upper integration limit
in case of bare $NN$ interactions.

\begin{figure}[!h]
\begin{center}
\includegraphics[width=0.49\textwidth] {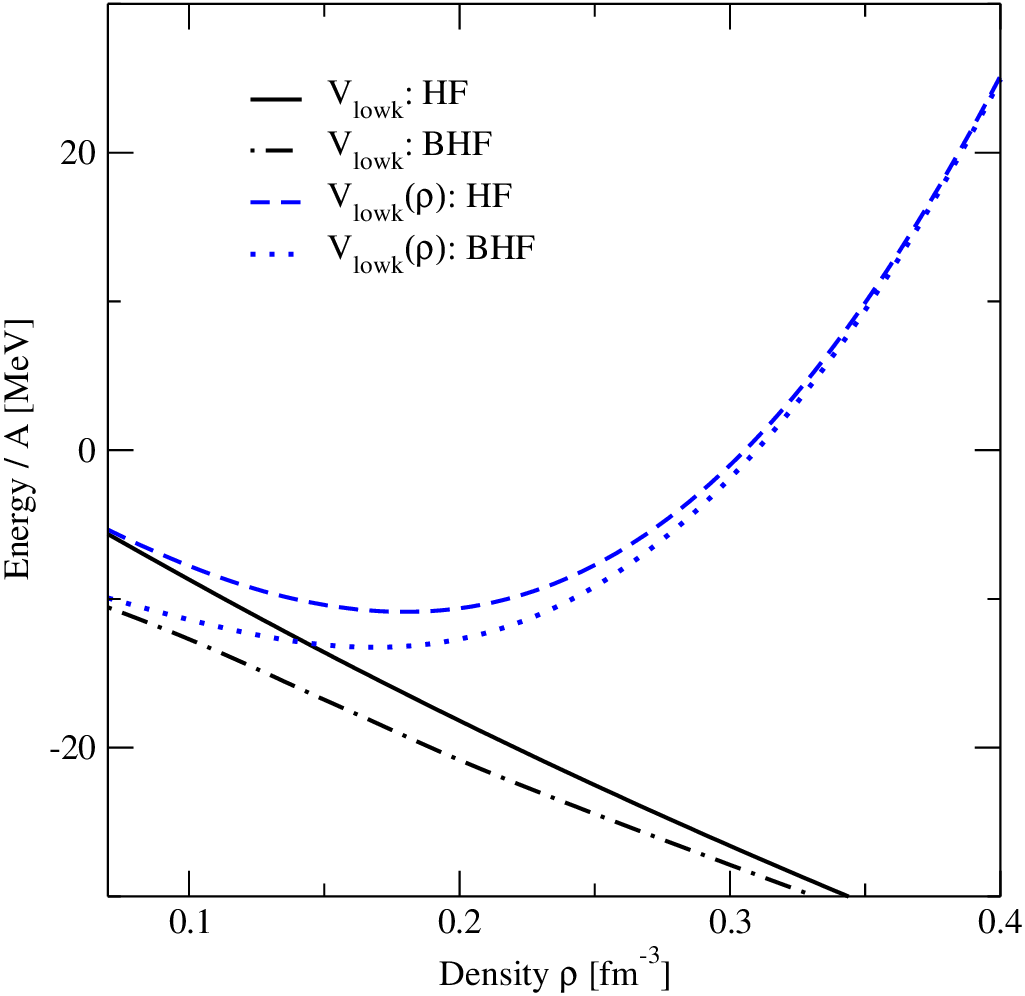}
\caption{Binding energy per nucleon of symmetric nuclear matter of a HF
calculation (dotted line) and of a BHF calculation (dashed line) employing a
density dependent $V_{lowk}(\rho)$ interaction. Furthermore, comparison is made
with a HF and a BHF calculation using a standard $V_{lowk}$ interaction, i.e. a
$V_{lowk}$ calculated at density $\rho = 0$. .
\label{fig:Vlowk}.}
\end{center}
\end{figure}
Results for the energy per nucleon of symmetric nuclear matter as a function of
the density $\rho$ obtained from HF and BHF calculations are displayed in 
Fig.~\ref{fig:Vlowk}. In all our calculations we use a cutoff $\Lambda$ of  2
fm$^{-1}$ and the Bonn A\cite{brockm} interaction has been employed for the
calculations displayed in this figure. The conventional method, which ignores
the medium modifications of the Dirac spinors, leads to results with the well
known features: The HF calculation using $V_{lowk}$ does not exhibit a minimum
in the energy as a function of the density. This absence of saturation is one of
the major problems of calculations for nuclear matter employing $V_{lowk}$. This
problem cannot be cured by the inclusion of correlations beyond the HF
approximation, e.g. by means of the BHF approximation.

These rather weak effects of the $NN$ correlations beyond mean-field can be
explained by the fact that only intermediate two-particle states with momenta
below the cutoff $\Lambda$ can be taken into account in Eq.~(\ref{eq:BG}), which
is not the case for bare $NN$ interactions. The $NN$ correlation effects even
vanish at high densities due to the lack of phase space for these correlations
and energies resulting from BHF calculations approach those determined in the HF
approximation.
 
The inclusion of relativistic effects by calculating the underlying Bonn A 
interaction
in
terms of dressed Dirac spinors, however,
drastically changes this picture.  The HF calculation for the density dependent
$V_{lowk}(\rho)$ interaction already leads to a minimum in the energy as a
function of the density.  Furthermore, this HF calculation yields a saturation
density in the neighborhood of the empirical region of saturation, but it yields
too little binding energy, i.e. the energy is about -11 MeV per nucleon. This
result for the binding energy can be improved by taking into account the effects
of $NN$ correlations within the BHF approximation. Hence, the BHF calculation
for the density dependent $V_{lowk}(\rho)$ interaction yields about 2 MeV more
binding in the region of saturation than the corresponding HF calculation due to
the additional correlations from $NN$ states. In short, the  mean-field
calculations using a density dependent $V_{lowk}(\rho)$ can already lead to
reasonable results and the $NN$ correlations beyond mean-field are rather weak,
in particular compared to mean-field calculations using bare $NN$
interactions~\cite{kuckei:2003}.

\begin{figure}[!h]
\begin{center}
\includegraphics[width=0.49\textwidth] {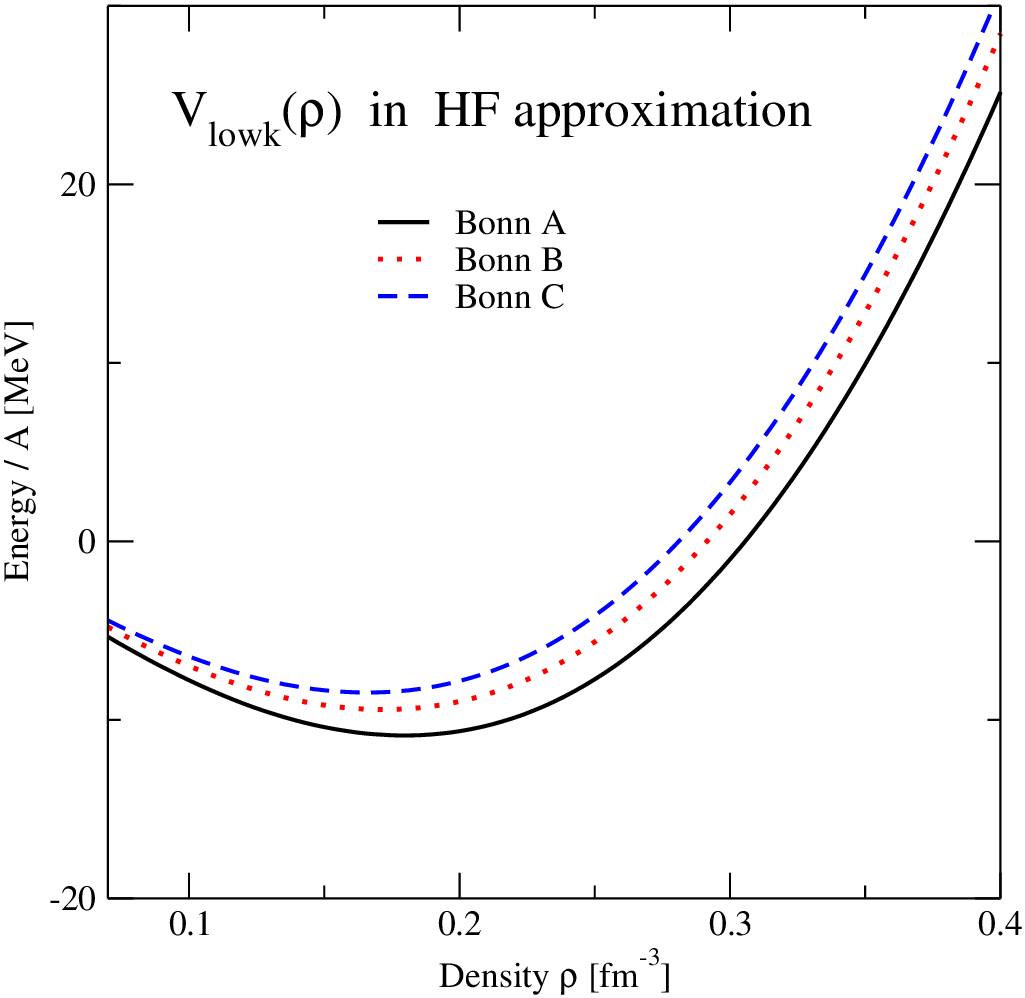}
\caption{Binding energy of nuclear matter as a function of density from
HF calculations employing various underlying potentials for the density dependent
effective interaction $V_{lowk}(\rho)$. \label{fig:HFBonnABC}}
\end{center}
\end{figure}

\begin{figure}[!h]
\begin{center}
\includegraphics[width=0.49\textwidth] {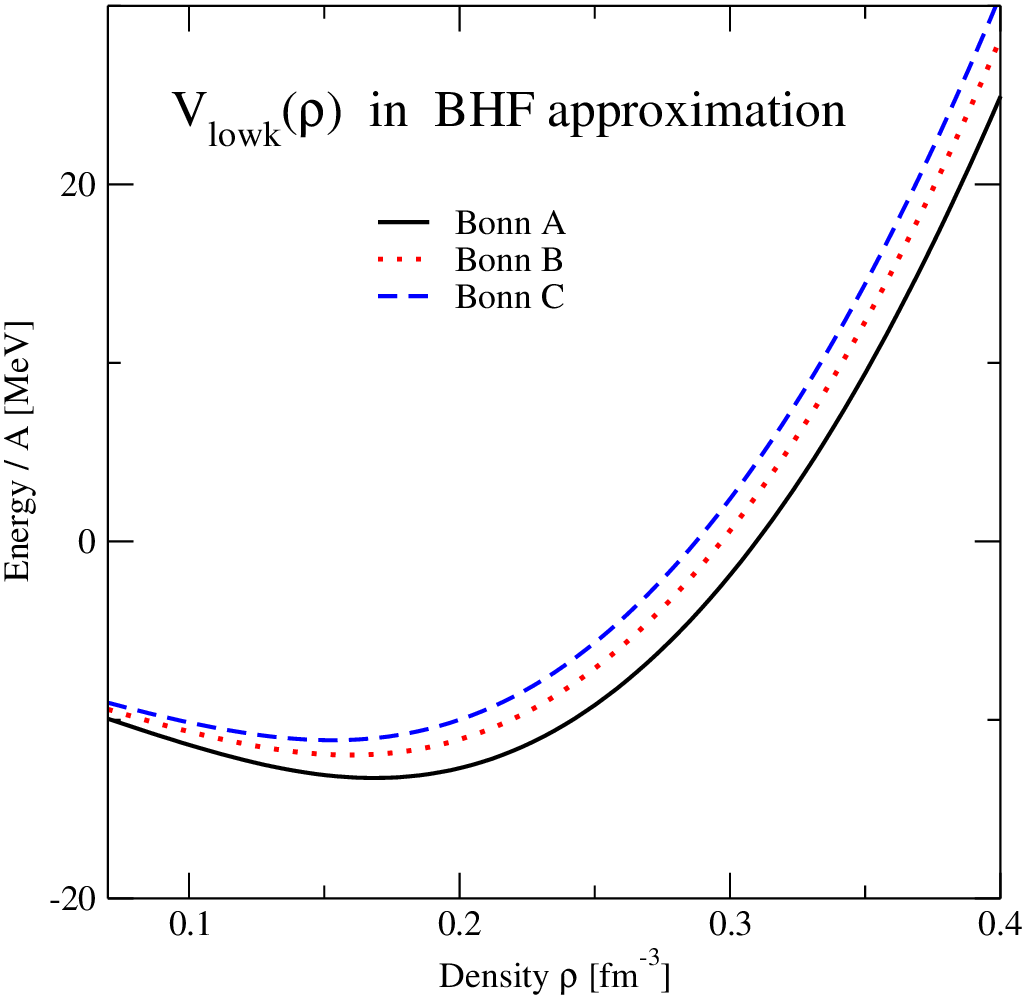}
\caption{Binding energy of nuclear matter as a function of density from
BHF calculations employing various underlying potentials for the density
dependent effective interaction $V_{lowk}(\rho)$. \label{fig:BHFBonnABC}}
\end{center}
\end{figure}

An attractive feature of the use of a standard $V_{lowk}$ interaction in HF or
BHF calculations  is that the corresponding results are rather insensitive to the
realistic interaction model on which the $V_{lowk}$ interaction is based.
Therefore, we want to explore to which extent this insensitivity applies in
nuclear matter calculations employing the density dependent $V_{lowk}(\rho)$
interaction. The HF calculations for density dependent $V_{lowk}(\rho)$
interactions based on different realistic interactions qualitatively shows the
same behavior, although quantitatively some difference in the binding energy
strength exists as can be seen in Fig.~\ref{fig:HFBonnABC}. The HF calculation 
using Bonn~A as underlying interaction yields about 2.2 MeV stronger binding in
the saturation point than the one using Bonn~C.  The same observations and
conclusions  about the sensitivity on the underlying realistic interaction can be
made for the BHF calculations in Fig.~\ref{fig:BHFBonnABC}. This dependence of
the calculated energy on the underlying $NN$ interaction is much weaker than the
corresponding difference of 8 MeV per nucleon obtained in non-relativistic BHF
calculations using directly Bonn A and C potentials\cite{brockm}. The
model-dependence of the relativistic $V_{lowk}$ interactions, however, seems to
be slightly stronger than the corresponding model-dependence in non-relativistic
calculations using the modern potentials, which fit the $NN$ data with high
precision. Here, one must keep in mind, however, that the potentials Bonn A, B
and C did not fit the phase-shifts that accurately and indeed a similar
model-dependence is also obtained applying the renormalisation scheme to these
potentials in the conventional way.

In summary, we study relativistic effects of nuclear matter calculations
employing an effective low momentum interaction. These effective low momentum
interactions are very interesting, since calculations with realistic bare $NN$
interactions are restricted to very light nuclei and not possible for heavier
nuclei or for the neutron star crust in the near future. However, one of the
major problems of these calculations employing an effective low momentum
interaction is the absence of saturation, unless three-body forces are added. 
Therefore, relativistic effects are introduced by dressing the Dirac spinors of
the underlying realistic interaction on which the low momentum interaction is
based. In this way a density dependent effective interaction $V_{lowk}(\rho)$ is
obtained. Employing this density dependent $V_{lowk}(\rho)$ interaction, one
already obtains a saturation point in the HF calculation.  The effects of the
$NN$ correlations introduced by the BHF approximation, although rather weak,
leads to a further improvement of the saturation properties. Hence, this leads to
the nice feature that HF calculations using a density dependent $V_{lowk}(\rho)$
can lead to reasonable results and $NN$ correlations beyond mean-field are rather
weak. This opens the door for studying finite nuclei in calculations which are
based on a realistic $NN$ interaction, treating correlation effects in a
perturbative manner.

This work has been supported by a grant (Mu 705/5-1) of 
the Deutsche Forschungsgemeinschaft (DFG).

\end{document}